\documentclass[runningheads,10pt]{llncs}
\usepackage{subfigure}

\usepackage{geometry}
\usepackage{amsfonts,amsmath,amssymb}

\usepackage{graphicx}
\usepackage{textcomp}
\usepackage{color}

\usepackage{soul}

\definecolor{darkred}{rgb}{0.5,0,0}
\definecolor{darkgreen}{rgb}{0,0.5,0}
\definecolor{darkblue}{rgb}{0,0,0.5}

\usepackage{url}
\usepackage{enumitem}
\setlist{noitemsep,leftmargin=*}
\setlist[itemize,1]{label=--}
\setlist[itemize,2]{label=-}
\newcommand{\prob}{\mathbf{Pr}}%
\newcommand{\advtage}{\mathbf{Adv}}%
\newcommand{\bdh}{BDH}%
\newcommand{\bimap}{e:G_1\times G_1\to G_2}%
\newcommand{\bilin}{e}%
\newcommand{\adva}{\mathcal{A}}%
\newcommand{\advb}{\mathcal{B}}%
\renewcommand{\S}{\mathcal{S}}%
\newcommand{\V}{\mathcal{V}}%
\newcommand{\usrf}{\mathcal{U}_1}%
\newcommand{\usrs}{\mathcal{U}_2}%
\newcommand{\id}{\mathsf{ID}}%
\newcommand{\ids}{\id_{\!\S}}%
\newcommand{\idv}{\id_{\!\V}}%
\newcommand{\G}{\mathbb{G}}%

\newcommand{\Z}{\mathbb{Z}}%

\newcommand{\setup}{\textsf{Setup}}%
\newcommand{\keygen}{\textsf{Key Extract}}%
\newcommand{\rgets}{\overset{\hbox{\tiny ~\$}}{\gets}}%
\newcommand{\gen}[1]{\langle #1 \rangle}%
\newcommand{\params}{\ensuremath{\mathit{params}}}%
\newcommand{\pub}{\ensuremath{P_{\!\!\mathit{pub}}}}%

\newcommand{\dvsig}{\textsf{DVBSig}}%{\sf{DVBSign}}%
\newcommand{\dvver}{\textsf{DVBVer}}%{\sf{DVBVer}}%{\sf{DVBVerify}}%
\newcommand{\dvsim}{\textsf{DVBSim}}%{\sf{DVBSim}}%{\sf{DVBTrans}}%

\newcommand{\bit}{b}%
\newcommand{\true}{1}%
\newcommand{\false}{0}%

\begin{document}
\mainmatter
\pagestyle{plain}

\title{Anonymous proof-of-asset transactions\\ using designated blind signatures}

\author{N. Sharma$^1$, R. Anand-Sahu$^2$, V. Saraswat$^3$, J. Garcia-Alfaro$^4$}

\institute{$^1$ Pt. Ravishankar Shukla University, India\\
$^2$ University of Luxembourg, Luxembourg.\\
$^3$ Robert Bosch Engineering \& Business Solutions Pvt. Ltd.,  India.\\
$^4$ Institut Polytechnique de Paris, T\'el\'ecom SudParis, France.
\vspace{-.5cm}
}

\maketitle

\begin{abstract}
We propose a scheme to preserve the anonymity of users in
proof-of-asset transactions. We assume bitcoin-like cryptocurrency
systems in which a user must prove the strength of its assets (i.e.,
solvency), prior conducting further transactions. The traditional way
of addressing such a problem is the use of blind signatures, i.e., a
kind of digital signature whose properties satisfy the anonymity of
the signer. Our work focuses on the use of a designated verifier
signature scheme that limits to only a single authorized party (within
a group of signature requesters) to verify the correctness of the
transaction. 
\end{abstract}

\section{Introduction}
\label{sec1}

Blind signature schemes offer a practical way of handling privacy
constraints in cryptocurrencry transactions~\cite{chaum1983blind}. A
blind signature construction is essentially an interactive two-party
protocol between the signer of a message and a group of signature
requesters. The signer disguises the contents of the message, before
signing it. This way, the signature requesters can verify the
correctness of the operation, without learning anything about the
message that has been signed. However, the validity of the signature
can be verified by anyone within the group of requesters. There may be
situations in which a particular signer wants to designate that just
one entity in the group of receivers must able to verify the signature
(and not the others). This is the objective of DVS (Designated
Verifier Signature)
schemes~\cite{jakobsson1996designated,saeednia2003efficient} addressed
in this paper.

In the realm of blockchain cryptocurrencies (i.e., bitcoin-like
digital cash schemes), the aforementioned situation may appear in the
so-called proof-of-asset transactions, in which users must prove their
solvency prior getting access to online services such as
cryptocurrency exchange markets. In other words, situations in which
users must prove that they control a given amount of assets (i.e.,
bitcoins) but without releasing the specific amount they owe. In
addition, we assume situations in which the users want to uniquely
designate who can verify those proof-of-asset transactions, e.g., to
avoid that a leakage of the proof is used by other parties (i.e.,
advertisement services, gambling platforms, etc.).

We address the aforementioned challenges and present a designated
verifier blind signature (DVS) construction using pairing-based
cryptography. The security of our scheme relies on the hardness of the
computational and the decisional bilinear Diffie-Hellman
problem~(cf. Section~\ref{app:preliminaries}). We analyze the security
of the approach and perform an efficiency comparison w.r.t. other
existing similar approaches.

\medskip

\noindent \textbf{Paper Organization ---} Section~\ref{sec:related}
surveys related work. Section~\ref{sec-background} provides some
preliminaries. Section~\ref{sec-our} presents our construction and
discusses about the security and efficiency of our approach.
Section~\ref{sec-con} concludes the paper.

\section{Related Work}
\label{sec:related}

Following seminal work by Chaum~\cite{chaum1983blind},
Boldyreva~\cite{boldyreva2002efficient} demonstrated and formalized
the concept of blind signature schemes under the random oracle model
and the computational Diffie-Hellman assumptions. Work by Chow et
al.~\cite{chow2005two} proved, as well, \emph{unlinkability}
properties of blind signatures. Camenisch et
al.~\cite{camenisch2004efficient} proposed new constructions without
random oracle constraints, without achieving proofs against strong
unforgeability. Liao et al.~\cite{liao2005pairing} provide new schemes
under the hardness of strong Diffie-Hellman assumptions. Zhang and
Kim~\cite{zhang2002id}, followed by Huang et
al.~\cite{huang2005efficient}, proposed identity-based blind
signatures, achieving unlinkability. Zhang et
al.~\cite{zhang2006linkability} uncovered linkability attacks
in~\cite{huang2005efficient} (signers being traced back under valid
message-signature pairs). Pointcheval and Stern~\cite{PS00} settled
fundamental security properties of blind signatures. Schr{\"o}der et
al.~\cite{schroder2012security} offer fair guidelines for the security
of blind signatures. They revisit the definition of unforgeability
in~\cite{PS00} and propose a new unforgeability definition to avoid
adversaries repeating a message for more than one signature.

The public verifiability of a signature is undesirable when a
signature shares sensitive information between the signer and the
verifier. To deal with this situation, the signer requires to sign the
document for a fixed receiver with control on its verification. For
this purpose, the idea of undeniable signature~\cite{CV90} was
suggested by Chaum and Van Antwerpen. Desmedt and Yung reported
in~\cite{desmedt1991weaknesses} some weaknesses in the aforementioned
approach. Jakobsson et al.~\cite{jakobsson1996designated} proposed a
non-interactive designated verifier proof which enables the signer to
produce transfer-resistant signatures for a designated verifier. In
other words, the verifier does not possess the capability to transfer
the proof of origin of the signature to third parties. Jakobsson et
al.~\cite{jakobsson1996designated} also suggested the necessity of
keeping the anonymity of signers. A concrete construction satisfying
such constraints (i.e., impossibility of transfer to third parties and
signer anonymity) was provided by Saeednia et al.
in~\cite{saeednia2003efficient}. Identity-based versions inspired by
the previous approach were presented by Susilo et al. in~\cite{SZM04},
and later by Zhang and Wen in~\cite{ZW07}. Limitations in~\cite{ZW07}
include the lack of proofs for unverifiability, non-transferability
and strongness, and the possibility of a signer with direct access to
the original message to blind and unblind messages and signatures,
hence not fulfilling the standard definition settled
in~\cite{chaum1983blind,PS00,schroder2012security}. The construction
presented in this paper addresses such shortcomings.

Bitcoin-like transaction anonymity has been addressed by Yi et al.
proposing schemes achieving blindness and
unforgeability~\cite{yi2019new}. More recently, Wang et al.
in~\cite{wang2020designated} has proposed the application of
designated verifier blind signatures for bitcoin proof-of-asset
transactions. When a vendor requires to an anonymous buyer to provide
a proof of solvency prior enabling an online service (e.g., a certain
amount of bitcoins), the buyer provides a proof about it in designated
manner. Hence, only the specific vendor requesting solvency to the
user can process the signature. The vendor cannot further use this
proof with any other third party. Our new construction addresses the
same problem, offering a more compact construction over pairings,
improving the efficiency of the identity-based construction by Zhang
and Wen in~\cite{ZW07}, and satisfying unverifiability,
non-transferability and strongness properties (cf.
Section~\ref{sec-background} and citations thereof).

\section{Preliminaries}
\label{sec-background}

\subsection{Identity-Based Cryptography}
\label{app:preliminaries}

\noindent A probabilistic polynomial time (PPT) algorithm is a
probabilistic random algorithm that runs in time polynomial in the
length of input. $y \rgets A(x)$ denotes a randomized algorithm $A(x)$
with input $x$ and output $y$. For $X$ being a set $v \rgets X$ stands
for a random selection of $v$ from $X$. A function $f: N \to [0,1]$ is
said to be negligible in $n$ if for any polynomial $p$ and for
sufficiently large $n$, the relation $f(n) < 1/p(n)$ holds. For an
element $g\in G$, where $G$ is a set, we denote the group $G=\gen{g}$
if $g$ generates or spans $G$.

\begin{definition}[Bilinear Map]\label{bilinear-map}{\textnormal{
Let $G_{1}$ and $G_{2}$ be two cyclic groups with a prime order $q$,
where $G_{1}$ is additive and $G_{2}$ is multiplicative. Let $P$ be
the generator of $G_{1}$. Then a map $\bimap$ is said to be a
cryptographic bilinear map if it fulfils the below conditions.
\begin{description}
\item[Bilinearity: ]
For all integers $x, y \in \Z_{q}^{*}$, $e(xA, yA) = e(A,A)^{xy}$, or
equivalently, for all $A,B,C \in G_{1}$, $e(A+B,C)=e(A, C)e(B, C)$ and
$e(A, B+C)=e(A, B)e(A, C)$.
\item[Non-Degeneracy: ] The points $A, B \in G_{1}$ with $e(A, B)\neq
  1$. As $G_{1}$ and $G_{2}$ are prime ordered groups this property is
  equivalent to have $g:=e(A,A) \neq 1$, or in other words $g:=e(A,A)$
  is a generator of $G_{2}$.
\item[Computability: ] The map $e(A, B)\in G_2$ can be computes
  efficiently for all $A,B \in G_{1}$.
\end{description}
}}
\end{definition}

\begin{definition}[Bilinear Map Parameter Generator]
{\textnormal{
A bilinear map parameter generator $\mathfrak{B}$ is a
PPT algorithm that takes as input security parameter $\lambda$
and outputs a tuple
\begin{equation}\label{bmpg}
\langle q,\bimap, P, g \rangle \gets \mathfrak{B}(\lambda)
\end{equation}
where $q$, $G_{1}$, $G_{2}$, $e$, $P$ and $g$ are as in
Definition~\ref{bilinear-map}. }}\end{definition}
\begin{definition}[Bilinear Diffie-Hellman Problem]
\label{bdhp}{\textnormal{
Given a security parameter $\lambda$, let
\[
\langle q,\bilin, P, g \rangle \gets \mathfrak{B}(\lambda) \,.\]
Let $\bdh: G_1 \times G_1 \times G_1 \to G_2$ be a map defined by
$\bdh(X, Y, Z) = \omega$ where
\[
X = xP, Y = yP, Z = zP \text{
and } \omega = e(P, P)^{xyz} \,.
\]
The bilinear Diffie-Hellman problem (BDHP) is to evaluate
$\bdh(X, Y, Z)$ given $X,Y,Z \rgets G_1$. (Without the knowledge of
$x,y,z\in\Z_q$ --- obtaining $x\in\Z_q$, given $P,X\in\G_1$ is
solving the discrete logarithm problem (DLP)).
}}\end{definition}
\begin{definition}[BDHP Parameter Generator]
{\textnormal{
A BDHP parameter generator $\mathfrak{C}$ is a PPT
algorithm that takes as input security parameter $\lambda$ and
outputs a tuple
\begin{equation}\label{bdhpg}
\langle q,\bimap, P, g, X,Y,Z \rangle \gets \mathfrak{C}(\lambda)
\end{equation}
where $q$, $G_{1}$, $G_{2}$, $e$, $P$, $g$, $X$, $Y$ and $Z$ are as
in Definition~\ref{bdhp}. }}\end{definition}
\begin{definition}[Bilinear Diffie-Hellman Assumption]
\label{bdha}{\textnormal{
Given a security parameter $\lambda$, let
\[
\langle q, \bilin, P,g, X,Y,Z \rangle \gets \mathfrak{C}(\lambda) \,.\].
The bilinear Diffie-Hellman assumption (BDHA) states that for any PPT
algorithm $\adva$ which attempts to solve BDHP, its
advantage $\advtage_{\mathfrak{C}}(\lambda)$, defined as
\[
\prob[\adva(q, \bilin, P,g, X,Y,Z) = \bdh(X,Y,Z)] \,,
\]
is negligible in $\lambda$.
}}\end{definition}
\begin{definition}[Decisional BDHP]
\label{dbdhp}
{\textnormal{
Given a security parameter $\lambda$, let
\[
\langle q, \bilin, P,g, X,Y,Z \rangle \gets \mathfrak{C}(\lambda) \,.\]
Let $\omega \rgets G_2$.
The decisional bilinear Diffie-Hellman problem (DBDHP) is to decide if
\[
\omega = \bdh(X, Y, Z) \,.
\]
That is, if $X = xP, Y = yP, Z = zP$, for some $x,y,z\in\Z_q$, then
the DBDHP is to decide if
\[
\omega = e(P, P)^{xyz} \,.
\]
(Without the knowledge of $x,y,z\in\Z_q$ --- obtaining $x\in\Z_q$,
given $P,X\in\G_1$ is solving the discrete logarithm problem (DLP)).
}}
\end{definition}
\begin{definition}[DBDHP Parameter Generator]
{\textnormal{
A DBDHP parameter generator $\mathfrak{D}$ is a PPT algorithm that
takes as input security parameter $\lambda$ and outputs a tuple
\begin{equation}\label{dbdhpg}
\langle q,\bimap, P, g, X,Y,Z,\omega \rangle \gets \mathfrak{D}(\lambda)
\end{equation}
where $q$, $G_{1}$, $G_{2}$, $e$, $P$, $g$, $X$, $Y$, $Z$ and
$\omega$ are as in Definition~\ref{dbdhp}. }}\end{definition}
\begin{definition}[Decisional BDHA]
\label{dbdha}{\textnormal{
Given a security parameter $\lambda$, let
\[
\langle q, \bilin, P,g, X,Y,Z, \omega \rangle \gets
\mathfrak{D}(\lambda) \,.\]
The bilinear Diffie-Hellman assumption (DBDHA) states that,
for any PPT algorithm $\adva$ which attempts to solve DBDHP, its
advantage $\advtage_{\mathfrak{D}}(\lambda)$, defined as
\begin{equation}
\left[~\!\!\!\begin{array}{l}
\prob[adva(q, \bilin, P,g, X,Y,Z,\omega)=1] \, - \\
\prob[\adva(q, \bilin, P,g, X,Y,Z,\bdh(X,Y,Z))=1]
\end{array}~\!\!\!\right],
\end{equation}
is negligible in $\lambda$.
}}\end{definition}

\subsection{Identity-based Strong Designated Verifier Blind Signatures}
\label{def3.2}
In this section, we provide formal definitions related with the construction of
identity-based strong designated verifier blind signature
(hereinafter, ID-SDVBS) schemes~\cite{ZW07}. In such schemes, a signer
with identity $\id_{S}$ intends to send a signed message to a
designated verifier with identity $\id_{V}$ such that no one other
than the designated verifier can verify the signature. The scheme
consists of the five algorithms described next:

\begin{enumerate}
\item $\params \gets \setup(\lambda)$: Executed by the Private Key
  Generator~(PKG), taking a security parameter $\lambda$ as input and
  producing, as output, the master secret $s$ and the public
  parameters ($\params$) of the system. The remaining algorithms
  listed below receive all the values of $\params$ as implicit inputs.

\item $(Q_{ID}, S_{ID})\gets \keygen(ID)$: The PKG takes as input an
  identity $\id$ and produces, as output, a public and private key
  pair $(Q_{\id}, S_{\id})$.

\item $\sigma \gets \dvsig(S_{\id_{S}}, Q_{\id_{V}}, m)$: Signer and
  user run this interactive process. Inputs include the signer's
  public and secret key $(Q_{\ids},S_{\ids})$, the designated
  verifier's public key $Q_{\id_{V}}$ and a message $m$. Signer and
  user stop this process in polynomial time, producing either a
  signature $\sigma$ of $m$, or false (in case an error happens).

\item $\bit \gets \dvver(S_{\id_{V}}, Q_{\id_{S}}, m, \sigma)$: Run by
  the verifier, taking as inputs $S_{\id_{V}}$ (secret key of the
  verifier), $Q_{\id_{S}}$ (public key of the signer), a message $m$
  and a signature $\sigma$. It returns a bit $\bit$ which is $\true$
  if the signature is valid (otherwise, it returns $\false$ if the
  signature is invalid).

\item $\widehat{\sigma} \gets \dvsim(Q_{\id_{S}}, S_{\id_{V}}, m)$:
  Run by the verifier, it takes as inputs $S_{\id_{V}}$ (verifier's
  secret key), $Q_{\id_{S}}$ and $Q_{\id_{V}}$ (public keys of the
  signer and the designated verifier), and a message $m$. It generates
  a signature $\widehat{\sigma}$ as output.
\end{enumerate}

\noindent Next, we provide definitions about the properties we aim to satisfy.

\begin{definition}[Correctness]\label{def-correctness}{\textnormal{If
the signature $\sigma$ on a message $m$ is correctly computed by
a signer $\id_{S}$, then the designated verifier $\id_{V}$ must be
able to verify the correctness of the message-signature pair $(m,
\sigma)$. That is,
\[\prob\Big[\true \leftarrow \dvver\Big(\begin{array}{l}
S_{\id_{V}},Q_{\id_{S}},m,\\\dvsig(Q_{\id_{V}},S_{\id_{S}},m)
\end{array}
\Big)\Big] = 1 \,.\]%
}}\end{definition}

\begin{definition}[Unforgeability]\label{def-euf}{\textnormal{%
An ID-SDVBS scheme is said to be strong existential
unforgeable against adaptive chosen message and adaptive
chosen identities attack if for any security parameter
$\lambda$, no probabilistic polynomial time adversary
$\adva(\lambda,\linebreak[0]t,\linebreak[0]\varepsilon,\linebreak[0]q_{H_1},\linebreak[0]q_{H_2},\linebreak[0]q_{E},\linebreak[0]q_{S},\linebreak[0]q_{V})$,
which runs in time $t$, has a non-negligible advantage
\begin{align*}
\varepsilon
&:= \advtage_{\text{ID-SDVBS},\adva}^{\text{SEUF-CID2-CMA2}}(\lambda) \\
&:= \prob[\true \gets \dvver(S_{\id_{V^*}}, Q_{\id_{S^*}}, m^*, \sigma^* )]
\end{align*}
against the challenger $\advb$ in the following game:
\begin{description}%
\item[\textnormal{1. \emph{Setup}:}]%
The challenger $\advb$ generates the systems public
parameter $\params$ for security parameter $\lambda$.
\item[\textnormal{2. \emph{Query Phase}:}]%
\begin{itemize}%
\item%
The adversary $\adva$ may request upto $q_{H_1}$ hash
queries on its adaptively chosen identities and upto and
$q_{H_2}$ hash queries on its adaptively chosen messages
and obtain responses from $\advb$ acting as a random
oracle.
\item%
$\adva$ may request upto $q_{E}$ key extraction queries on
its adaptively chosen identities and obtain the
corresponding private keys.
\item%
$\adva$ may request upto $q_{S}$ signature queries on its
adaptively chosen messages and adaptively chosen
identities for the signer and the designated verifier and
obtain a valid strong designated verifier signature.
\item%
$\adva$ may request upto $q_{V}$ verification queries on
signatures on its adaptively chosen messages $m$ and
adaptively chosen identities for the signer and the
designated verifier and obtain the verification result
$\true$ if it is valid and $\false$ if invalid.
\end{itemize}%
\item[\textnormal{3. \emph{Output}:}]%
Finally, $\adva$ outputs a (message, signature) pair $(m^*,
\sigma^*)$ for identities $\id_{S}^*$ of the signer and
$\id_{V}^*$ of the designated verifier such that:
\begin{itemize}%
\item%
$\adva$ has never submitted $\id_{S}^*$ or $\id_{V}^*$
during the key extraction queries.
\item%
$\sigma^*$ was never given as a response to a signature
query on the message $m^*$ with the signer's identity
$\id_{S}^*$, and the designated verifier's identity
$\id_{V}^*$;
\item%
$\sigma^*$ is a valid signature on the message $m^*$ from
a signer with identity $\id_{S}^*$ for a designated
verifier with identity $\id_{V}^*$.
\end{itemize}%
\end{description}%
}}\end{definition}%

\begin{definition}[Unverifiability]\label{def-dver}{\textnormal{
An ID-SDVBS scheme is said to be existential designated unverifiabile
against adaptive chosen message and adaptive chosen identities attack
if for any security parameter $\lambda$, no probabilistic polynomial
time adversary
$\adva(\lambda,\linebreak[0]t,\linebreak[0]\varepsilon,\linebreak[0]q_{H_1},
\linebreak[0]q_{H_2},\linebreak[0]q_{E},\linebreak[0]q_{S},\linebreak[0]q_{V})$
which runs in time $t$ has a non-negligible advantage
\begin{align*}
\varepsilon
&:= \advtage_{\text{ID-SDVBS},\adva}^{\text{EDV-CID2-CMA2}}(\lambda) \\
&:= \big|\prob[\adva(Q_{\id_{S^*}}, Q_{\id_{V^*}}, m^*,\sigma^*)=1] \, - \\
&\!\!\prob[\adva(Q_{\id_{S^*}} \!,\! Q_{\id_{V^*}} \!,\!
m^* \!,\!\dvsig(S_{\id_{S^*}} \!,\! Q_{\id_{V^*}} \!,\! m^*)) \!=\! 1] \big|
\end{align*}
against the challenger $\advb$'s response $\sigma^*$ in the following
game:
\begin{description}
\item[\textnormal{1. \emph{Setup}:}] Challenger $\advb$ generates
  the system public parameters $\params$ from $\lambda$.
\item[\textnormal{2. \emph{Query Phase 1}:}]
\begin{itemize}
\item Adversary $\adva$ may request up to $q_{H_1}$ hash queries on
  its adaptively chosen identities; and up to $q_{H_2}$ hash queries
  on its adaptively chosen messages. $\adva$ may obtain responses from
  $\advb$, acting as a random oracle.
\item $\adva$ may request upto $q_{E}$ key extraction queries on its
  adaptively chosen identities and obtain the corresponding private
  keys.
\item $\adva$ may request upto $q_{S}$ signature queries on its
  adaptively chosen messages and adaptively chosen identities for the
  signer and the designated verifier and obtain a valid strong
  designated verifier signature.
\item $\adva$ may request upto $q_{V}$ verification queries on
  signatures on its adaptively chosen messages $m$ and adaptively
  chosen identities for the signer and the designated verifier and
  obtain the verification result $\true$ if it is valid and $\false$
  if invalid.
\end{itemize}
\item[\textnormal{3. \emph{Challenge}:}] At some point, $\adva$
  outputs a message $m^*$ and identitie $\id_{S}^*$ of the signer and
  $\id_{V}^*$ of the designated verifier on which it wishes to be
  challenged such that $\adva$ has never submitted $\id_{S}^*$ or
  $\id_{V}^*$ during the key extraction queries. The challenger
  $\advb$ responds with a ``signature'' $\sigma^*$ and challenges
  $\adva$ to verify if it is valid or not.
\item[\textnormal{4. \emph{Query Phase 2}:}] $\adva$ continues its
  queries as in Query Phase 1 with an additional restriction that now
  it cannot submit a verification query on $\sigma^*$.
\item[\textnormal{5. \emph{Output}:}] $\adva$ outputs a bit $\bit^*$
  which is $\true$ if the signature is valid and $\false$ if invalid.
\end{description}
}}\end{definition}

\begin{definition}[Non-transferability]
\label{def-trans}
{\textnormal{ An ID-SDVBS scheme is said to achieve
    non-transferability if the signature generated by the signer is
    computationally indistinguishable from that generated by the
    designated verifier, that is, {\small{
\[
\sigma \!\gets\! \dvsig(Q_{\id_{V}}, S_{\id_{S}}, m) ~\approx~
\widehat{\sigma} \!\gets\! \dvsim(Q_{\id_{S}}, S_{\id_{V}}, m) .
\]}}}}
\end{definition}

\begin{definition}[Strongness]
\label{def-strong}
{\textnormal{An ID-SDVBS scheme is said to be strong designated if
    given $\sigma \gets \dvsig(S_{\id_{S}}, Q_{\id_{V}}, m)$, anyone,
    say $V^*$, other than the designated verifier $V$ can produce
    identically distributed transcripts that are indistinguishable
    from those of $\sigma$ from someone, say $S^*$, except the signer
    $S$. That is, {\small{
\[
\sigma \!\gets\! \dvsig(Q_{\id_{V}}, S_{\id_{S}}, m) ~\approx~
\widehat{\sigma} \!\gets\! \dvsim(Q_{\id_{S^{*}}},
S_{\id_{V^{*}}}, m) .
\]}}}}
\end{definition}

\begin{definition}[Blindness]
\label{def-blindness}
{\textnormal{ An ID-SDVBS scheme must ensure the fact that the signer
    knows nothing about the message she signs. In other way, after
    producing signatures on different messages, the signer cannot
    relate that which message corresponds to which signature. More
    precisely, if signer has made a list of certain messages, and the
    requests for signatures have been placed by the user by picking
    messages randomly from that list, then after looking all the
    signatures together, the signer cannot make a list of
    corresponding (messages, signature) pairs. Our security model for
    the blindness is motivated by the models considered in
    \cite{JU97,zhang2002id}. Let $\adva$ be a probabilistic polynomial
    time adversary/algorithm which has control over the malicious
    signer. Let $\usrf$ and $\usrs$ be two honest users who interact
    with the signer in the following attack game.
\begin{enumerate}%
\item%
$\adva$ is provided responses to it's \emph{Key extraction queries}
$(Q_{ID}, S_{ID})\gets \keygen(ID)$ as in the security game for
unforgeability.
\item%
$\adva$ outputs
messages $m_0, m_1$.
\item%
$b\in\{0,1\}$ is defined to be a bit. User $\usrf$ and $\usrs$ randomly selects a bit $b$ and pick $m$ and $m'$ as their random input taps. Here $m$ is corresponding to bit $b$ and so as $m'$ corresponding to $(b-1)$.
\item%
$\adva$ communicates with users $\usrf$ and $\usrs$ in the random order during the signature issuing protocol.
\item%
If the user $\usrf$ does not fail and posses a signature $\sigma_m$ and also user $\usrs$ does
not fail and posses signature $\sigma_{m'}$, then $\adva$ is provided these additional information
$\sigma_m$ and $\sigma_{m'}$, which are outputs essentially based on the bit $b$ and $(b-1)$, the actual value of the bit depends upon the value of $b$ of-course. The game does not abort, and continue, even
if either of the users fails but the other does not, and the corresponding output is forwarded to
$\adva$ in that case.
\item%
Finally, $\adva$ outputs a bit $b'\in \{0,1\}$ which is $\true$ if $b' = b$.
\end{enumerate}%
We define advantage of the adversary $\adva$ in the above game, by following
\[
Adv^{Blind}_{\adva} = (2 \times Pr[b'=b]) - 1
\]
It is straightforward that $\adva$ can always output a true bit with probability $\frac{1}{2}$. But, in this case the advantage is clearly $0$. A signature is said to satisfying blindness, if there is no probabilistic polynomial-time algorithm/adversary
$\adva$ who wins the above game with non-negligible advantage.
}}\end{definition}

\section{Proposed Construction}
\label{sec-our}

Find below the algorithms of our construction ($\setup$,
$\keygen$, Designated Blind Signature $\dvsig$, Designated
Verification $\dvver$ and Transcript Simulation $\dvsim$):

\begin{itemize}
\item \textbf{$\setup$ --} In this algorithm, on input security
  parameter $\lambda$ PKG outputs the master private key $s \in
  \Z_{q}^{*}$ and the public parameters
  \[
  params = (1^{\lambda}, G_{1}, G_{2}, q, e, H_{1}, H_{2}, P, \pub)
  \,,
  \]
  where $G_{1}$ is an additive cyclic group of prime order $q$ with
  generator $P$, $G_{2}$ is a multiplicative cyclic group of prime
  order $q$, and $H_{1} : \{0, 1\}^{*} \to G_{1}$, $H_{2} : \{0,
  1\}^{*} \times G_{1} \to \Z_{q}^{*}$ are two cryptographic secure
  collision resistant hash functions, and $P_{pub} = sP \in G_{1}$ is
  system's public key, $\bimap$ is a bilinear map (cf.
  Section~\ref{app:preliminaries}).

\item \textbf{$\keygen$ --} On input identity $ID_{i} \in \{0,
  1\}^{*}$, the PKG computes public key as $Q_{ID_{i}} = H_{1}(ID_{i})
  \in G_{1}$ and private key as $S_{ID_{i}} =
  sQ_{ID_{i}} \in G_{1}$, for this identity.

\item \textbf{$\dvsig$ --} Signs message $m \in \{0,1\}^{*}$ (it can
  be verified by a designated verifier $V$).

  \begin{enumerate}
  \item Signer $S$ selects a random $r \rgets \Z_{q}^{*}$ and calculates:
    \begin{itemize}
    \item $U= rQ_{\ids}\in G_{1}$;
    \end{itemize}

  \item As commitment, the signer sends the calculated value $U$ to
    the user.

  \item \emph{Blinding Phase}: In this algorithm, user selects $x,y
    \in_{R} Z_{q}^{*}$ as blinding factors. Then calculates

    \begin{itemize}
    \item $U' = x U +  xy Q_{\ids}$;
    \item $h = H_{2}(m, U' ) \in \Z_{q}^{*}$;
    \item $h_{1} = x^{-1}h + y$;
    \end{itemize}

  \item User sends this calculated value $h_{1}$ to the signer.

  \item \emph{Signing Phase}: After receiving $h_{1}$, the signer calculates

    \begin{itemize}
    \item $V = (r+ h_{1})S_{\ids}\in G_{1}$;
      and sends back the user the value $V$.
    \end{itemize}

  \item \emph{Unblinding Phase}: Then, the user calculates
    \begin{itemize}
    \item $V' = x V$

    \item $\sigma = e(V', Q_{\idv})$. $(U', \sigma) \in G_{1} \times
      G_{2}$ is the strong designated verifier blind signature on the
      message $m$.
    \end{itemize}
  \end{enumerate}

\item \textbf{$\dvver$ --} After receiving signature $(U', \sigma)$ on
  a  message $m$, a verifier computes $h = H_{2}(m, U') \in \Z_{q}^{*}$
  and accepts the signature if $ \sigma = e( U' + hQ_{\ids}, S_{\idv})$.

\item \textbf{$\dvsim$ --} The designated verifier $V$ can produce the
  same signature $\widehat{\sigma}$ intended for itself, by performing
  this algorithm: chooses an integer $\widehat{r},\widehat{x},
  \widehat{y} \rgets \Z_{q}^{*}$ and computes:%
  \begin{itemize}
  \item $\widehat{U}= \widehat{r}Q_{\ids} \in G_{1}$;
  \item $\widehat{U'}= \widehat{x}U + \widehat{x} \widehat{y} Q_{\ids} \in G_{1}$;
  \item $\widehat{h} = H_{2}(m, \widehat{U'}) \in \Z_{q}^{*}$;
  \item $\widehat{h_{1}} = \widehat{x}^{-1} \widehat{h} + \widehat{y}$;
  \item $\widehat{V} = (\widehat{r} + \widehat{h_{1}})Q_{\ids}\in G_{1}$;
  \item $\widehat{V'} = \widehat{x} \widehat{V}$; and
  \item $\widehat{\sigma} = e(\widehat{V'}, S_{\idv})$.
  \end{itemize}
\end{itemize}

\subsection{Security Analysis}
\label{sec-security}

The verification of our proposed scheme is as follows.
If the generated signature $(U',\sigma)$ on a message $m$ from a
signer with identity $\ids$ for a designated verifier with identity
$\idv$, then the proposed scheme follows:%
\begin{align*}
\label{correctness}
  e(U' + hQ_{\ids}, S_{\idv})
    &= e(x U + xyQ_{\ids}+ hQ_{\ids}, S_{\idv})
  \\&= e(x U + xy Q_{\ids}+ hQ_{\ids}, sQ_{\idv})
  \\&= e(x rQ_{\ids} + xy Q_{\ids}+ hQ_{\ids},sQ_{\idv})
  \\&= e(x( rQ_{\ids} +yQ_{\ids}+  x^{-1}hQ_{\ids}), sQ_{\idv})
  \\&= e(x( rS_{\ids} +y S_{\ids}+ x^{-1}hS_{\ids}),Q_{\idv})
  \\&= e(x(r +(x^{-1}h+ y ))S_{\ids}, Q_{\idv})
  \\&= e(x(r + h_{1})S_{\ids}, Q_{\idv})
  \\&= e(x V, Q_{\idv})
  \\&= e(V', Q_{\idv})
  \\&= \sigma~.
\end{align*}

\noindent Next, we discuss the achievement of the following security properties:
(1) unforgeability, (2) unverifiability, (3) non-transferability, (4) strongness and (5) blindness.

\begin{theorem}(Unforgeability)
  \label{thmUnforgeability}
  Given a security parameter $\lambda$, if there exists a PPT
  adversary
  $\adva(\lambda,\linebreak[0]t,\linebreak[0]\varepsilon,\linebreak[0]q_{H_1},\linebreak[0]q_{H_2},\linebreak[0]q_{E},\linebreak[0]q_{S},\linebreak[0]q_{V})$
  which breaks the unforgeability of the proposed ID-SDVBS scheme in
  time $t$ with success probability $\varepsilon$,
  then there exists a PPT adversary
  $\advb(\lambda,t',\varepsilon')$ which solves BDHP with
  success probability at least
  \begin{align*}%\label{}
    \varepsilon' \geq \Big(& 1 -\frac{1}{q^2} \Big)
    \Big( 1 -\frac{2}{q_{H_{1}}} \Big)^{q_{E}  + q_{V}} \\&%
    \Big( 1 -\frac{2}{q_{H_{1}}(q_{H_{1}}  - 1)} \Big)^{q_{S}}%
    \Big( \frac{2}{q_{H_{1}}(q_{H_{1}}  - 1)} \Big)\varepsilon
\end{align*}
  in time at most
\begin{align*}%\label{}
      t' \leq &(q_{H_1}  + q_E  + 3q_S  + q_V) S_{G_1}  \\&%
      + (q_S  + q_V) P_{e}  +
      q_S O_{G_1}  + O_{G_2}  + S_{G_2}  + t
\end{align*}
  where $S_{G_1}$ (resp. $S_{G_2}$) is the time taken for one scalar
  multiplication in $G_1$ (resp. $G_2$), $O_{G_1}$ (resp. $O_{G_2}$)
  is the time taken for one group operation in $G_1$ (resp. $G_2$),
  and $P_e$ is the time taken for one pairing computation.
\end{theorem}%

\paragraph{\textbf{Proof of Theorem~\ref{thmUnforgeability}:}}%
Let for a security parameter $\lambda$, $\advb$ is challenged to solve
the BDHP for
\[%
\langle q,e,G_1,G_2,P,aP,bP,cP \rangle
\]%
where $G_1$ is an additive cyclic group of prime order $q$ with
generator $P$, $G_2$ is a multiplicative cyclic group of prime order
$q$ with generator $e(P,P)$, and $\bimap$ is a
cryptographic bilinear map (cf. Section~\ref{app:preliminaries}).
$a,b,c \rgets \Z_q^*$ are unknown to $\advb$. The goal of $\advb$ is
to solve BDHP by computing $e(P,P)^{abc} \in G_2$ using $\adva$, the
adversary who claims to forge our proposed ID-SDVBS scheme. $\advb$
simulates the security game for unfogeability with $\adva$ as follows.

\begin{description}%
\item[\textnormal{\emph{Setup}:}]%
  $\advb$ generates the systems public parameter
  \[%
  \params = \langle q,\bimap,P,\pub:=cP,H_1,H_2
  \rangle %
  \]%
  for security parameter $\lambda$ where the hash functions $H_1$ and
  $H_2$ behave as random oracles and responds to $\adva$'s queries as
  below.

\item[\textnormal{\emph{$H_{1}$-queries}:}]%
  To respond to the $H_1$ queries, $\advb$ maintains a list
  \[L_{H_1}=\{( \id_i \in \{0,1\}^*, r_i \in \Z_q^*, R_i \in G_1
  )_{i=1}^{q_{H_1}}\}\]%
  which is initially empty.
  $\advb$ randomly chooses two indices $\alpha,\beta\in[1,q_{H_1}]$
  and sets $i=0$. When $\adva$ makes an
  $H_{1}$-query for an identity $\id \in \{0,1\}^{*}$
  $\advb$ proceeds as follows.
  \begin{enumerate}%
  \item%
    If the query $\id$ already appears in $L_{H_{1}}$ in some tuple
    $(\id_i, r_i, R_i)$ then $\advb$ responds to $\adva$ with
    $H_{1}(\id) = R_i \in G_1$;
  \item%
    otherwise $\advb$ sets $i=i+1$ and
    \begin{itemize}%
    \item%
      if $i = \alpha$, $\advb$ sets $r_i = \perp$ and $R_i = aP$;
    \item%
      if $i = \beta$, $\advb$ sets $r_i = \perp$ and $R_i = bP$;
    \item%
      if $i\neq\alpha,\beta$, $\advb$ chooses $r_{i} \rgets
      \Z_{q}^{*}$ and sets $R_i = r_i P$;
    \end{itemize}%
  \item%
    Finally, $\advb$ adds the tuple $(\id_i:=\id, r_i, R_i)$ to
    $L_{H_1}$ and responds to $\adva$ with $H_{1}(\id) = R_{i}$.
  \end{enumerate}%

\item[\textnormal{\emph{$H_{2}$-queries}:}]%
  For response of $H_2$ queries, $\advb$ maintains a list
  \[L_{H_2}=\{( (m,U') \in \{0,1\}^* \times G_1, h \in \Z_q^* )
  \}\]%
  initially which is empty.  When $\adva$ queries the oracle $H_{2}$
  at $(m, U')$, $\advb$ responds as follows.
  \begin{enumerate}%
  \item%
    If the query $(m, U')$ already appears in the $H_{2}$-list in
    the tuple $(m, U', h)$ then $\advb$ respond with $H_{2}(m,
    U') = h \in \Z_{q}^{*}$.
  \item%
    Otherwise $\advb$ picks a random $h \in \Z_{q}^{*}$ and adds the
    tuple $(m, U', h)$ to the $H_{2}$-list and responds to $\adva$
    with $H_{2}(m, U') = h$.
  \end{enumerate}%

\item[\textnormal{\emph{Key extraction queries}:}]%
  When $\adva$ makes a private key query on identity $\id$, $\advb$
  proceeds as follows.
  \begin{enumerate}%
  \item%
    Runs the above algorithm for responding to $H_{1}$-query for
    identity $\id$ and obtains $H_{1}(\id) = R_i$.
  \item%
    If $i=\alpha\text{ or }\beta$, $\advb$ reports failure and halts.
  \item%
    If $i\neq\alpha,\beta$, $\advb$ responds to $\adva$ with the
    private key $S_{\id} := r_{i} \pub$ on the identity $\id$.
  \end{enumerate}%

  It can be verified that the provided private key $S_{\id} =
  r_{i}\pub$ is a valid private key for the user with identity
  $\id_{i}:=\id$ since
  \[%
  r_{i}\pub = r_{i} cP = c r_{i} P = c H_{1}(\id) \,.
  \]%

  Note that $\advb$ aborts the security game during a key extraction
  query with probability $\frac{2}{q_{H_{1}}}$.

\item[\textnormal{\emph{Signature queries}:}]%
  To respond to the signature queries, $\advb$ maintains a list
  \begin{align*}%\label{}
    L_{S}=\{(&
    m_\ell \in \{0,1\}^*, {\ids}_{\ell} \in \{0,1\}^*, {\idv}_{\ell} \in \{0,1\}^*, \\&%
    x_\ell \in \Z_q^*, U'_\ell \in G_1, \sigma_\ell \in G_2
    )_{\ell=1}^{q_{S}}\}
\end{align*}
  which is initially empty with $\ell=0$.
  When $\adva$ queries the signature on a message $m$ from a signer
  with identity $\ids$ for a designated verifier with identity $\idv$,
  $\advb$ proceeds as follows.
  \begin{enumerate}%
  \item%
    If the query $(m,\ids,\idv)$ already appears in $L_{S}$ in some
    tuple
    $(m_\ell,\linebreak[0]{\ids}_{\ell},\linebreak[0]{\idv}_{\ell},\linebreak[0]x_\ell,\linebreak[0]U'_\ell,\linebreak[0]\sigma_\ell)$
    then $\advb$ responds to $\adva$ with the signature $(U'_\ell,
    \sigma_\ell)$.
  \item%
    Otherwise $\advb$ sets $\ell=\ell+1$ and for responding to $H_{1}$-query for identities $\ids$ and $\idv$
runs the above algorithm and obtains $H_{1}(\ids) = R_i$ and $H_{1}(\idv) = R_j$.
  \item%
    If $\{i,j\}=\{\alpha,\beta\}$, $\advb$ reports failure and halts.
  \item%
    If $i\neq\alpha,\beta$, then $\advb$ computes the private key for
    $\ids$, $S_{\ids} = r_{i}\pub$ and proceeds as follows.
    \begin{itemize}%
    \item%
      randomly chooses $x_{\ell} \in \Z_{q}^{*}$;
    \item%
      sets $U'_{\ell}= x_{\ell}P \in G_{1}$;
    \item%
      runs the $H_{2}$-query algorithm to obtain $h_{\ell} = H_{2}(m,
      U'_{\ell}) \in \Z_{q}^{*}$;
    \item%
      sets $V'_{\ell} = x_{\ell}\pub + h_{\ell} S_{\ids} \in
      G_{1}$;
    \item%
      computes $\sigma_{\ell} = e(V'_{\ell}, Q_{\id_{j}}=R_j)$.
    \end{itemize}%

  \item%
    Otherwise if $j\neq\alpha,\beta$, then $\advb$ computes the
    private key $\idv$, $S_{\idv} = r_{j}\pub$ and proceeds as
    follows.
    \begin{itemize}%
    \item%
      randomly chooses $x_{\ell} \in \Z_{q}^{*}$;
    \item%
      sets $U'_{\ell}= x_{\ell}P \in G_{1}$;
    \item%
      runs the $H_{2}$-query algorithm to obtain $h_{\ell} = H_{2}(m,
      U'_{\ell}) \in \Z_{q}^{*}$;
    \item%
      sets $V'_{\ell} = x_{\ell}\pub + h_{\ell} S_{\idv} \in
      G_{1}$;
    \item%
      computes $\sigma_{\ell} = e(V'_{\ell}, Q_{\id_{i}}=R_i)$.
    \end{itemize}%

  \item%
    Finally, $\advb$ adds the tuple $(m_\ell, {\ids}_{\ell},
    {\idv}_{\ell}, x_\ell, U'_\ell, \sigma_\ell)$ to $L_{S}$ and
    responds to $\adva$ with the signature $(U'_\ell,
    \sigma_\ell)$.
  \end{enumerate}%
  Note that $\advb$ aborts the security game during a signature query
  with probability $\frac{2}{q_{H_1}(q_{H_1}-1)}$.
\item[\textnormal{\emph{Verification queries}:}]%
  When $\adva$ makes a verification query on the signature
  $(U',\sigma)$ on a message $m$ from a signer with identity $\ids$
  for a designated verifier with identity $\idv$, $\advb$ proceeds as
  follows.
  \begin{enumerate}%
  \item%
    $\advb$ runs the above algorithm for responding to $H_{1}$-query
    for identities $\ids$ and $\idv$ and obtains $H_{1}(\ids) = R_i$
    and $H_{1}(\idv) = R_j$.
  \item%
    If $j\in\{\alpha,\beta\}$, $\advb$ reports failure and halts.
  \item%
    If $j\neq\alpha,\beta$, then $\advb$ computes $\idv$'s private
    key, $S_{\idv} = r_{j}\pub$, and proceeds as in the verification
    of the proposed scheme and responds to $\adva$ accordingly.
  \end{enumerate}%

  Note that $\advb$ aborts the security game during a verification
  query with probability
  $\frac{2}{q_{H_1}}$.

\item[\textnormal{\emph{Output}:}]%
  After $\adva$ has made its queries, it finally outputs a valid
  signature $(U^{'*}, \sigma^{*})$ on a message $m^*$ from a signer
  with identity $\ids^{*}$ for a designated verifier with identity
  $\idv^{*}$ with a non-negligible probability
  $\varepsilon$ such that:
  \begin{itemize}%
  \item%
    $\adva$ has never submitted $\ids^*$ or $\idv^*$ during the key
    extraction queries;
  \item%
    $(U^{'*}, \sigma^{*})$ was never given as a response to a
    signature query on the message $m^*$ with the signer's identity
    $\ids^*$, and the designated verifier's identity $\idv^*$; and
  \item%
    $\sigma^{*} = e(U^{'*} + h^{*}Q_{\ids}, S_{\idv})$.
  \end{itemize}%
\end{description}%

If $\adva$ did not make $H_1$-query for the identities $\ids^{*}$ and
$\idv^{*}$, then the probability that verification equality holds is
less than $1/q^2$. Thus, with probability greater than $1-1/q^2$, both
the public keys were computed using $H_1$-oracle and there exist
indices $i,j \in [1,q_{H_1}]$ such that $\ids^{*}=\id_{i}$ and
$\idv^{*}=\id_{j}$. If $\{i,j\} \neq \{\alpha,\beta\}$, then $\advb$
reports failure and terminates.

\paragraph{Solution to BDHP: }%
Otherwise, as in the forking lemma, $\advb$ repeats the game with the
same random tape for $x_{\ell}$ but with different choices of a random
set for $H_{2}$-queries
to obtain another forgery $(U^{*}, \sigma')$ on the message $m^{*}$
with $h'$ such that $h^{*} \neq h'$ and $\sigma^{*} \neq \sigma'$.
Then,
\begin{align}\label{forgery-eqn}%
  \frac{\sigma^{*}}{\sigma'} &= \frac{e(U^{'*} + h^{*}Q_{\ids},
    S_{\idv})}{e(U^{'*} + h'Q_{\ids}, S_{\idv})} \nonumber\\&=
  \frac{e(h^{*}Q_{\ids}, S_{\idv})}{e(h'Q_{\ids}, S_{\idv})} \nonumber\\&=
  \frac{e(Q_{\ids}, S_{\idv})^{h^{*}}}{e(Q_{\ids}, S_{\idv})^{h'}}\\
  %\nonumber\\
&= e(Q_{\ids}, S_{\idv})^{(h^{*} - h')} \nonumber\\&= e(aP, bcP)^{(h^{*} - h')}%
  \nonumber\\&= (e(P, P)^{abc})^{(h^{*} - h')} \,.
\end{align}%

Let $(h^{*} - h')^{-1} \mod{q} = \hat{h}$. Then, from the above
equation, $\advb$ solves the BDHP by computing
\begin{equation}\label{forgery-cdhp}%
  e(P,P)^{abc} = %\Big( \frac{\sigma^{*}}{\sigma'} \Big)^{\hat{h}}%
  (\sigma^{*}/\sigma')^{\hat{h}}%
\end{equation}%

\paragraph{Probability calculation: }%
If $\advb$ does not abort during the simulation then $\adva$'s view is
identical to its view in the real attack. The responses to
$H_{1}$-queries and $H_{2}$-queries are as in the real attack, since
each response is uniformly and independently distributed in $G_{1}$
and $\Z_{q}^{*}$ respectively. The key extraction, signature and
verification queries are answered as in the real attack.

The probability that $\advb$ does not abort during the simulation is
\begin{equation}\label{forge-prob-abort}%
  \Big(1-\frac{2}{q_{H_{1}}}\Big)^{q_{E} + q_{V}}%
  \Big(1-\frac{2}{q_{H_{1}}(q_{H_{1}} - 1)}\Big)^{q_{S}}%
  % \Big(\frac{2}{q_{H_{1}}(q_{H_{1}}- 1)}\Big)
  \,.\end{equation}%

The probability that $\adva$ did $H_1$-query for the identities
$\ids^{*}$ and $\idv^{*}$ and that
$\{\ids^{*},\idv^{*}\}=\{\id_{\alpha},\id_{\beta}\}$ is
\begin{equation}\label{forge-prob-h1}%
  \Big(1-\frac{1}{q^2}\Big)
  \Big(\frac{2}{q_{H_{1}}(q_{H_{1}}- 1)}\Big) \,.
\end{equation}%

Clearly $\advb$'s advantage $\varepsilon'$ for solving
the BDHP, that is, the total probability that $\advb$ succeeds to
solve BDHP, is the product of $\adva$'s advantage
$\varepsilon$ of forging the proposed ID-SDVBS and
the above two probabilities. Hence
\begin{align*}%\label{}
\varepsilon' \geq \Big(& 1-\frac{1}{q^2}\Big)
\Big(1-\frac{2}{q_{H_{1}}}\Big)^{q_{E} + q_{V}} \\&%
\Big(1-\frac{2}{q_{H_{1}}(q_{H_{1}} - 1)}\Big)^{q_{S}}%
\Big(\frac{2}{q_{H_{1}}(q_{H_{1}} -
  1)}\Big)\varepsilon \,.
\end{align*}

\paragraph{Time calculation: }%
It can be observed that running time of the algorithm $\advb$ is same
as that of $\adva$ plus time taken to respond to the hash queries, key
extraction queries, signature queries and verification queries,
$q_{H_1}+q_{H_2}+q_{E}+q_{S}+q_{V}$. Hence the maximum running time
required by $\advb$ to solve the BDHP is
\begin{align*}%\label{}
  t' \leq & (q_{H_1} + q_E + 3q_S + q_V) S_{G_1} + (q_S + q_V) P_{e} \\&%
  + q_S O_{G_1} + O_{G_2} + S_{G_2} + t
\end{align*}
as $\advb$ requires to compute one scalar multiplication in $G_1$ to
respond to $H_1$ hash query, one scalar multiplication in $G_1$ to
respond to key extraction query, three scalar multiplications in $G_1$
to respond to signature query, one scalar multiplication in $G_1$ to
respond to verification query; one pairing computation to respond to
signature query, one pairing computation to respond to verification
query, one group operation in $G_1$ to respond to signature query,
and, one group operation in $G_2$ and one scalar multiplication in
$G_2$ to output a solution of BDHP. \qed

\medskip

\begin{theorem}(Unverifiability)
  \label{thmUnverifiability}
  Given a security parameter $\lambda$, if there exists a PPT
  adversary
  $\adva(\lambda,\linebreak[0]t,\linebreak[0]\varepsilon,\linebreak[0]q_{H_1},
  \linebreak[0]q_{H_2},\linebreak[0]q_{E},\linebreak[0]q_{S},\linebreak[0]q_{V})$
  which breaks the designated unverifiability of the proposed ID-SDVBS
  scheme in time $t$ with success probability $\varepsilon$, then
  there exists a PPT adversary $\advb(\lambda,t',\varepsilon')$ which
  solves DBDHP with success probability at least
  \begin{align*}
    \varepsilon' \geq &\Big(1-\frac{1}{q^2}\Big)
    \Big(1-\frac{2}{q_{H_{1}}}\Big)^{q_{E} + q_{V}}\\&
    \Big(1-\frac{2}{q_{H_{1}}(q_{H_{1}} - 1)}\Big)^{q_{S}}
    \Big(\frac{2}{q_{H_{1}}(q_{H_{1}} -
      1)}\Big)\varepsilon
  \end{align*}
  in time at most
  \begin{align*}
    t' \leq &(q_{H_1} + q_E + 3q_S + q_V) S_{G_1} + (q_S + q_V) P_{e} \\&
    + q_S O_{G_1} + S_{G_1} + S_{G_2} + P_{e}+t
  \end{align*}
  where $S_{G_1}$ (resp. $S_{G_2}$) is the time taken for one scalar
  multiplication in $G_1$ (resp. $G_2$), $O_{G_1}$ (resp. $O_{G_2}$)
  is the time taken for one group operation in $G_1$ (resp. $G_2$),
  and $P_e$ is the time taken for one pairing computation.
\end{theorem}

\paragraph{\textbf{Proof of Theorem \ref{thmUnverifiability}:}}%
Let for a security parameter $\lambda$, $\advb$ is challenged to solve
the DBDHP for
\[%
\langle q,\bimap,P,aP,bP,cP,\omega \rangle %
\]%
where $G_1$ is an additive cyclic group of prime order $q$ with
generator $P$, $G_2$ is a multiplicative cyclic group of prime order
$q$ with generator $e(P,P)$, and $\bimap$ is a
cryptographic bilinear map as described in Section~\ref{sec-prelims}%
and $\omega \rgets G_2$. $a,b,c \rgets \Z_q^*$ are unknown to $\advb$.
The goal of $\advb$ is to solve DBDHP by verifying if
$e(P,P)^{abc}=\omega$ using $\adva$, the adversary who claims to forge
our proposed ID-SDVBS scheme.

$\advb$ simulates the security game for strongness with $\adva$ by
doing the Setup and by responding the $H_{1}$-queries,
$H_{2}$-queries, Key extraction queries, Signature
  queries and Verification queries as in the security game for
unforgeability.

\paragraph{Output: }%
After $\adva$ has made its queries, it finally outputs a
% valid signature $(U^{'*}, \sigma^*)$ on
message $m^*$, an identity $\ids^{*}$ of a signer and an identity
$\idv^{*}$ of a designated verifier on which it wishes to be
challenged.

If $\adva$ did not make $H_1$-query for the identities $\ids^{*}$ and
$\idv^{*}$, then the probability that verification equality holds is
less than $1/q^2$. Thus, with probability greater than $1-1/q^2$, both
the public keys were computed using $H_1$-oracle and there exist
indices $i,j \in [1,q_{H_1}]$ such that $\ids^{*}=\id_{i}$ and
$\idv^{*}=\id_{j}$. If $\{i,j\} \neq \{\alpha,\beta\}$, then $\advb$
reports failure and terminates.

\paragraph{Solution to DBDHP: }%
Otherwise, $\advb$
\begin{itemize}%
\item%
  chooses a random $r \rgets \Z_q^*$;
\item%
  sets $U' = rP$;
\item%
  sets $h = H_2(m^{*},U')$;
\item%
  sets $\sigma = e(bP,cP)^{r} \omega^{h}$;
\end{itemize}%
and challenges $\adva$ to verify the validity of the signature
$(U,\sigma)$.
Then, the verification holds if and only if each of the following
holds
\begin{align*}&&%
  \sigma &= e(U' + h Q_{\ids} , S_{\idv}) \\%
  \Longleftrightarrow &&%
  \sigma &= e(rP + h aP , b\pub) \\%
  \Longleftrightarrow &&%
  \sigma &= e(rP , b\pub) e(h aP , b\pub) \\%
  \Longleftrightarrow &&%
  \sigma &= e(P , b\pub)^{r} e(aP , b\pub)^{h} \\%
  \Longleftrightarrow && %
  \sigma &= e(bP , \pub)^{r} e(aP , b\pub)^{h} \\%
  \Longleftrightarrow &&%
  \sigma &= e(bP , cP)^{r} e(aP , bcP)^{h} \\%
  \Longleftrightarrow &&%
  \sigma &= e(bP , cP)^{r} (e(P , P)^{abc})^{h} \\%
  \Longleftrightarrow &&%
  e(bP,cP)^{r} \omega^{h} &= e(bP , cP)^{r} (e(P , P)^{abc})^{h} \\%
  \Longleftrightarrow && %
  \omega^{h} &= (e(P , P)^{abc})^{h} \\%
  \Longleftrightarrow &&%
  \omega &= e(P , P)^{abc} \label{forgery-eqn}%
\end{align*}%

Then, from the above equation, $\advb$ solves the DBDHP by simply
returning the response of $\adva$ to the strongness challenge.

\paragraph{Probability calculation: }%
If $\advb$ does not abort during the simulation then $\adva$'s view is
identical to its view in the real attack. The responses to
$H_{1}$-queries and $H_{2}$-queries are as in the real attack, since
each response is uniformly and independently distributed in $G_{1}$
and $\Z_{q}^{*}$ respectively. The key extraction, signature and
verification queries are answered as in the real attack.

The probability that $\advb$ does not abort during the simulation is
\begin{equation}\label{ver-prob-abort}%
  \Big(1-\frac{2}{q_{H_{1}}}\Big)^{q_{E} + q_{V}}%
  \Big(1-\frac{2}{q_{H_{1}}(q_{H_{1}} - 1)}\Big)^{q_{S}}%
  % \Big(\frac{2}{q_{H_{1}}(q_{H_{1}}- 1)}\Big)
  \,.\end{equation}%

The probability that $\adva$ did $H_1$-query for the identities
$\ids^{*}$ and $\idv^{*}$ and that $\ids^{*}=\id_{\alpha}$ and
$\idv^{*}=\id_{\beta}$ is
\begin{equation}\label{ver-prob-h1}%
  \Big(1-\frac{1}{q^2}\Big)
  \Big(\frac{2}{q_{H_{1}}(q_{H_{1}}- 1)}\Big) \,.
\end{equation}%

Clearly $\advb$'s advantage $\varepsilon'$ for solving
the DBDHP, that is, the total probability that $\advb$ succeeds to
solve DBDHP, is the product of $\adva$'s advantage
$\varepsilon$ of breaking the strongness of the
proposed ID-SDVBS and the above two probabilities. Hence
\begin{align*}%\label{}
\varepsilon' \geq &\Big(1-\frac{1}{q^2}\Big)
\Big(1-\frac{2}{q_{H_{1}}}\Big)^{q_{E} + q_{V}}\\&%
\Big(1-\frac{2}{q_{H_{1}}(q_{H_{1}} - 1)}\Big)^{q_{S}}%
\Big(\frac{2}{q_{H_{1}}(q_{H_{1}} -
  1)}\Big)\varepsilon \,.
\end{align*}

\paragraph{Time calculation: }%
It can be observed that running time of the algorithm $\advb$ is same
as that of $\adva$ plus time taken to respond to the hash queries, key
extraction queries, %signature,
signature queries and verification queries, that is,
$q_{H_1}+q_{H_2}+q_{E}+q_{S}+q_{V}$.  Hence the maximum running time
required by $\advb$ to solve the BDHP is
\begin{align*}%\label{}
  t' \leq &(q_{H_1} + q_E + 3q_S + q_V) S_{G_1} + (q_S + q_V) P_{e} \\&%
  + q_S O_{G_1} + S_{G_1} + S_{G_2} + P_{e}+t
\end{align*}
as $\advb$ requires to compute one scalar multiplication in $G_1$ to
respond to $H_1$ hash query, one scalar multiplication in $G_1$ to
respond to key extraction query, three scalar multiplications in $G_1$
to respond to signature query, one scalar multiplication in $G_1$ to
respond to verification query; one pairing computation to respond to
signature query, one pairing computation to respond to verification
query, one group operation in $G_1$ to respond to signature query,
and, one scalar multiplication in $G_1$, one scalar multiplication in
$G_2$ and one pairing computation to output a solution of DBDHP. \qed

\begin{theorem}(Non-transferability)
\label{thmNon-transferability}
Let $\sigma$ be the signature generated by the signer. Then, $\sigma$ is
computationally indistinguishable, i.e., {\small{
\[
\sigma \!\gets\! \dvsig(Q_{\id_{V}}, S_{\id_{S}}, m) ~\approx~
\widehat{\sigma} \!\gets\! \dvsim(Q_{\id_{S}}, S_{\id_{V}}, m) .
\]}}
\end{theorem}

\paragraph{\textbf{Proof of Theorem \ref{thmNon-transferability}}:}
  As defined in Section \ref{def3.2}, the property \emph{non-transferability} holds when the signatures generated by the signer is indistinguishable from the one generated by the designated verifier. To show this property for our proposed scheme, below we show that: two signatures $(U', \sigma)$ and $(\widehat{U'},
  \widehat{\sigma})$, generated on message $m$ by
  the signer and the designated verifier respectively, are indistinguishable. It can be observed that the indistinguishability holds immediately as the two distributions:
\begin{center}\begin{minipage}[m]{0.4\columnwidth}\noindent
 \begin{align*}
   U &= rQ_{\ids};\\[0.45ex]
   U'&= x U + xy Q_{\ids};\\[0.45ex]
   h &= H_{2}(m, U');\\[0.45ex]
   h_{1} &= x^{-1}h + y;\\[0.45ex]
   V &= (r + h_{1})S_{ID_{S}};\\[0.45ex]
   V' &= x V;\\[0.45ex]
   \sigma &= e(V', Q_{ID_{V}});
 \end{align*}
   \end{minipage}\quad\begin{minipage}[m]{0.1\columnwidth}\noindent
     and
   \end{minipage}\begin{minipage}[m]{0.4\columnwidth}\noindent
 \begin{align*}
 \widehat{U} &=  \widehat{r}Q_{\ids};\\
 \widehat{U'} &=  \widehat{x}U + xy
       Q_{\ids};\\
 \widehat{h}  &=  H_{2}(m, \widehat{U'});\\
 \widehat{h_{1}}  &=  \widehat{x}^{-1} h +
       \widehat{y};\\
 \widehat{V}  &=  (\widehat{r} + \widehat{h_{1}})Q_{\ids};\\
 \widehat{V'}  &=  \widehat{x}V;\\
 \widehat{\sigma}  &=  e(\widehat{V'}, S_{\idv});
 \end{align*}
   \end{minipage}\end{center}
are identical. \qed

\begin{theorem}(Strongness)
\label{thmStrongness}
Let $\sigma$ be the signature generated by a signer $S$.
Let $V$ be the designated verifier, such
that $\sigma \gets \dvsig(S_{\id_{S}}, Q_{\id_{V}}, m)$.
Then, only $V$ can produce identically distributed
transcripts that are indistinguishable from those
of $\sigma$ from someone, say $S^*$, except the
signer $S$. That is, {\small{
\[
\sigma \!\gets\! \dvsig(Q_{\id_{V}}, S_{\id_{S}}, m) ~\approx~
\widehat{\sigma} \!\gets\! \dvsim(Q_{\id_{S^{*}}},
S_{\id_{V^{*}}}, m) .
\]}}
\end{theorem}

\paragraph{\textbf{Proof of Theorem \ref{thmStrongness}:}}
  By using the description below, it can be evidenced that our
  ID-SDBVS scheme achieves \emph{strongness} as defined in Section
  \ref{def3.2}. For this purpose we essentially show that for a
  signature $\sigma \gets \dvsig(Q_{\idv}, S_{\ids}, m)$ generated by
  the signer $\S$ for the designated verifier $\V$, the designated
  verifier $\V^*$ (other than $\V$) can generate a signature $\sigma
  \gets \dvsim(Q_{\ids^{*}}, S_{\idv^{*}}, m)$, as a signature
  generated by the signer $\S^*$ (other than $\S$) using the
  transcript simulation, where $Q_{\ids^{*}}$ and $S_{\idv^{*}}$ are
  defined as in the following, since
  \begin{align*}
    \sigma &=  e(x rS_{\ids} +x h_{1}S_{\ids}, Q_{\idv})\\
    &= e(x rS_{\ids} + x h_{1}S_{\ids}, mQ_{\idv^*})
    &&\text{\!\!\!\!\!where } Q_{\idv} \!\!=\! mQ_{\idv^{*}}\\
    &= e(x rmS_{\ids} + x h_{1}mS_{\ids}, Q_{\idv^*}) \\
    &= e(x rS_{\ids^{*}} + x h_{1}S_{\ids^{*}}, Q_{\idv^{*}})
    &&\text{\!\!\!\!\!where }S_{\ids^{*}} \!\!=\!  mS_{\ids}.\\
    &= e(x rQ_{\ids^{*}} + x h_{1}Q_{\ids^{*}}, S_{\idv^{*}})
  \end{align*}
\qed

\begin{theorem}(Blindness)
  \label{thmBlindness}
Let $x,y \in_{R} Z_{q}^{*}$ denote a random selection of
blinding factors. Let $\advb$ simulate the security game for blindness with $\adva$.
Let $\adva$ be the probabilistic polynomial-time algorithm and has $(Q_{ID}, S_{ID})$ from the key extraction queries as in the security game for unforgeability. Then, our proposed signature scheme satisfies the blindness property established in Definition~\ref{def-blindness}.
\end{theorem}

\paragraph{\textbf{Proof of Theorem \ref{thmBlindness}:}}%

We follow the technique from \cite{zhang2002id} to prove
Theorem \ref{thmBlindness}. If two signatures $(U'_i, \sigma'_i)$ and
$(U'_j, \sigma'_j)$ generated by users  $\mathcal{U}_i(U_i, h_{1_i}, V_i)$ and $\mathcal{V}_i(U_j, h_{1_j}, V_j)$for ($i,j\in \{0,1\}$) are provided to adversary $\adva$ (who is in control over the signer, but not over the users), the adversary cannot draw the true bit $b$ in a correct order, corresponding to the received signatures, where
$(U_i, h_{1_i}, V_i)$ and $(U_j, h_{1_j}, V_j)$ are essentially the values exchanged between the users and the signer during the interactive signature protocol.

It is sufficient to show that there exist two random factors $(X', y')$  that maps $(U_i, h_{1_i}, V_i)$ to $(U'_j, \sigma'_j)$ for each $i, j \in \{0,1\}$~(where $\sigma'_j=e(V'_j, Q_{\idv})$ and $X' \in G_{1}$). We define $X' =V_i-  V'_j $ (where $V'_j =  r'S_{\ids} + h'_{1_j}S_{\ids}; r' \in_{R} Z_{q}^{*}$ ) and $y' =  - h_{1_i} - (-h'_{1_j})$, since:

\begin{align*}%\label{}%
  \sigma & = e(V_i, Q_{\idv}) \\&%
  = e(X'+V'_j, Q_{\idv}) \\&%
  = e(X' + r'S_{\ids} + h'_{1_j}S_{\ids}, Q_{\idv}) \\&%
  = e(X', Q_{\idv})e( r'Q_{\ids} + h'_{1_j}Q_{\ids}, S_{\idv}) \\&%
  = e(X', Q_{\idv})e( U'_j + h'_{1_j}Q_{\ids}, S_{\idv}) \\&%
  = e(X', Q_{\idv})e( U'_j,S_{\idv}) e( (y' + h_{1_i})Q_{\ids}, S_{\idv}) \\&%
  = e(X', Q_{\idv})e(U'_j+ (h_{1_i}+ y')Q_{\ids},S_{\idv})  \,.
\end{align*}%

Hence, we  can claim that there will always exist random values, i.e.,
the blinding factors, which hold the same relation as in the signature
issuing protocol.\qed

\subsection{Performance Estimation}

\noindent Inspired by the performance analysis discussed by Debiao et
al. in~\cite{DJJ11}, we discuss next the expected computation time for
the generation and verification of signatures using our approach.

We assume the same pairing used by Debiao et al., i.e., a Tate
pairing, which is capable of achieving an equivalent of 1024-bit RSA
security. It is defined over the supersingular elliptic curve $E=F_p :
y^2=x^3+x$ with embedding degree $2$ was used, where $q$ is a 160-bit
Solinas prime $q=2^{159}+2^{17}+1$ and $p$ a 512-bit prime satisfying
$p+ 1=12qr$. Accordingly, operation times are assumed as follows:
$6.38$ ms for each scalar multiplication; $5.31$ ms for each
exponentiation in $G_2$; $3.04$ ms for each map-to-point hash
execution; and $20.04$ ms for each pairing computation. Other
operations, such as the cost of an inverse operation over $Z_{q}^{*}$,
are omitted in our analysis, since it takes less than $0.03$ ms.
Likewise, the operation time of performing one general hash function
is also omitted, since it is expected to take less than $0.001$ ms,
hence negligible compared to the time taken by aforementioned (most
costly) operations (cf.~\cite{DJJ11} and citations thereof for further
details).

A careful analysis of our approach shows that each signature
generation would require five scalar multiplications (i.e., $6.38$ ms
each), one map-to-point hash execution (i.e., $3.04$ ms), and one
pairing computation (i.e., $20.04$ ms). In other words, our approach
would require about $54.98$ ms per signature generation. In terms of
signature verification, our approach would require one scalar
multiplication, one map-to-point hash execution and one pairing
computation. Hence, leading to about $29.46$ ms per signature
verification. If we conduct now the same analysis to the closest
approach in the literature, i.e., the identity-based construction by
Zhang and Wen in~\cite{ZW07}, we would obtain about $67.74$ ms per
signature generation (i.e., five scalar multiplications, one
map-to-point hash execution and one pairing computation) and $89.58$
ms per signature verification (i.e., one scalar multiplications, one
map-to-point hash execution and four pairing computations). Hence, and
by using the performance analysis in~\cite{DJJ11}, our construction
offers higher efficiency while addressing the limitations
in~\cite{ZW07} (i.e., lack of \emph{Blinding} and \emph{Unblinding}
procedures in their signature protocol, as well as lack of
unverifiability, non-transferability and strongness properties).

\section{Conclusion}
\label{sec-con}

We have presented a designated verifier signature scheme to enable
anonymity in proof-of-asset transactions. It allows cryptocurrency
users to prove their solvency in a privacy-friendly manner, while
designating a single authorized party (from a group of signature
requesters) to be able to verify the correctness of the transaction.
The approach uses pairing-based cryptography. More precisely, an
adaptive approach using an identity-based setting. The security of our
construction has been proved using the hardness assumption of the
decisional and computational bilinear Diffie-Hellman problem. We have
also presented an early estimation of the computation cost of our
approach, in terms of signature generation and signature verification.
The estimation shows that the computational cost and operation time of
the new scheme is significantly more efficient that previous efforts
in the literature, while addressing the previous limitations.

\bibliographystyle{abbrv}
\bibliography{references}

\begin{thebibliography}{10}

\bibitem{boldyreva2002efficient}
A.~Boldyreva.
\newblock Efficient threshold signature, multisignature and blind signature
  schemes based on the gap-diffie-hellman-group signature scheme.
\newblock {\em IACR ePrints}, 2002:118, 2002.

\bibitem{camenisch2004efficient}
J.~Camenisch, M.~Koprowski, and B.~Warinschi.
\newblock Efficient blind signatures without random oracles.
\newblock In {\em ICSCN}, pages 134--148. Springer, 2004.

\bibitem{chaum1983blind}
D.~Chaum.
\newblock Blind signatures for untraceable payments.
\newblock In {\em Advances in cryptology}, pages 199--203. Springer, 1983.

\bibitem{CV90}
D.~Chaum and H.~Van~Antwerpen.
\newblock Undeniable signatures.
\newblock In {\em Conference on the Theory and Application of Cryptology},
  pages 212--216. Springer, 1989.

\bibitem{chow2005two}
S.~S. Chow, L.~C. Hui, S.-M. Yiu, and K.~Chow.
\newblock Two improved partially blind signature schemes from bilinear
  pairings.
\newblock In {\em ACISP}, pages 316--328. Springer, 2005.

\bibitem{DJJ11}
H.~Debiao, C.~Jianhua, and H.~Jin.
\newblock An id-based proxy signature schemes without bilinear pairings.
\newblock {\em Annals of Telecommunications}, 66(11-12):657--662, 2011.

\bibitem{desmedt1991weaknesses}
Y.~Desmedt and M.~Yung.
\newblock Weaknesses of undeniable signature schemes.
\newblock In {\em Theory and Application of of Cryptographic Techniques}, pages
  205--220. Springer, 1991.

\bibitem{huang2005efficient}
Z.~Huang, K.~Chen, and Y.~Wang.
\newblock Efficient identity-based signatures and blind signatures.
\newblock In {\em ICCNS}, pages 120--133. Springer, 2005.

\bibitem{jakobsson1996designated}
M.~Jakobsson, K.~Sako, and R.~Impagliazzo.
\newblock Designated verifier proofs and their applications.
\newblock In {\em International Conference on the Theory and Applications of
  Cryptographic Techniques}, pages 143--154. Springer, 1996.

\bibitem{JU97}
A.~Juels, M.~Luby, and R.~Ostrovsky.
\newblock Security of blind digital signatures.
\newblock {\em Advances in Cryptology?CRYPTO'97}, pages 150--164, 1997.

\bibitem{liao2005pairing}
J.~Liao, Y.~Qi, P.~Huang, and M.~Rong.
\newblock Pairing-based provable blind signature scheme without random oracles.
\newblock In {\em ICCIS}, pages 161--166. Springer, 2005.

\bibitem{PS00}
D.~Pointcheval and J.~Stern.
\newblock Security arguments for digital signatures and blind signatures.
\newblock {\em Journal of cryptology}, 13(3):361--396, 2000.

\bibitem{saeednia2003efficient}
S.~Saeednia, S.~Kremer, and O.~Markowitch.
\newblock An efficient strong designated verifier signature scheme.
\newblock In {\em ICISC}, pages 40--54. Springer, 2003.

\bibitem{schroder2012security}
D.~Schr{\"o}der and D.~Unruh.
\newblock Security of blind signatures revisited.
\newblock In {\em International Workshop on Public Key Cryptography}, pages
  662--679. Springer, 2012.

\bibitem{SZM04}
W.~Susilo, F.~Zhang, and Y.~Mu.
\newblock Identity-based strong designated verifier signature schemes.
\newblock In {\em Australasian Conference on Information Security and Privacy},
  pages 313--324. Springer, 2004.

\bibitem{wang2020designated}
H.~Wang, D.~He, and Y.~Ji.
\newblock Designated-verifier proof of assets for bitcoin exchange using
  elliptic curve cryptography.
\newblock {\em Future Generation Computer Systems}, 107:854--862, 2020.

\bibitem{yi2019new}
X.~Yi and K.-Y. Lam.
\newblock A new blind ecdsa scheme for bitcoin transaction anonymity.
\newblock In {\em Proceedings of the 2019 ACM Asia Conference on Computer and
  Communications Security}, pages 613--620, 2019.

\bibitem{zhang2002id}
F.~Zhang and K.~Kim.
\newblock Id-based blind signature and ring signature from pairings.
\newblock In {\em International Conference on the Theory and Application of
  Cryptology and Information Security}, pages 533--547. Springer, 2002.

\bibitem{zhang2006linkability}
J.~Zhang, T.~Wei, J.~Zhang, and W.~Zou.
\newblock Linkability of a blind signature scheme and its improved scheme.
\newblock In {\em International Conference on Computational Science and Its
  Applications}, pages 262--270. Springer, 2006.

\bibitem{ZW07}
N.~Zhang and Q.~Wen.
\newblock {Provably Secure Blind ID-Based Strong Designated Verifier Signature
  Scheme}.
\newblock In {\em CHINACOM'07}, pages 323--327. IEEE, 2007.

\end{thebibliography}

\end{document}